\documentclass[prd,superscriptaddress,twocolumn,letterpaper,%
nofootinbib,showpacs,preprintnumbers,amsmath,amssymb]{revtex4}
\usepackage{bm,curves}
\newcommand{\del}{\partial}
\newcommand{\gaz}{g_A^{\mbox{$\scriptscriptstyle (Z)$}}}
\newcommand{\run}[1]{\widetilde{\alpha}_{#1}}
\newcommand{\hsp}[1]{\hspace*{#1 mm}}
\newcommand{\smallfrac}[2]{\mbox{\small ${\displaystyle \frac{#1}{#2}}$}}
\newcommand{\footfrac}[2]%
{\mbox{\footnotesize ${\displaystyle \frac{#1}{#2}}$}}

\begin{document}

\title{Heavy-quark axial charges to non-leading order}
\author{S.D. Bass}
\affiliation{Institut f\"ur Theoretische Physik, Universit\"at Innsbruck,
Technikerstrasse 25, A 6020 Innsbruck, Austria}
\author{R.J. Crewther}
\affiliation{Department of Physics and Mathematical Physics and \linebreak
Centre for the Subatomic Structure of Matter (CSSM), 
University of Adelaide, SA 5005, Australia}
\affiliation{Centre for Particle Theory, Department of Mathematical
Sciences, University of Durham, South Rd, Durham DH1 3LE, United
Kingdom}
\author{F.M. Steffens}
\affiliation{Instituto de Fisica Teorica,
Rua Pamplona 145, 01405-900 Sao Paulo - SP, Brazil}
\author{A.W. Thomas}
\affiliation{Department of Physics and Mathematical Physics and \linebreak
Centre for the Subatomic Structure of Matter (CSSM), 
University of Adelaide, SA 5005, Australia}

\begin{abstract}
We combine Witten's renormalization group with the matching conditions 
of Bernreuther and Wetzel to calculate at next-to-leading order
the complete heavy-quark contribution to the neutral-current axial-charge 
measurable in neutrino-proton elastic scattering.  Our results are
manifestly renormalization group invariant.
\end{abstract}
\preprint{ADP-02-77/T516}
\maketitle

\section{Introduction}

This paper announces results for the next-to-leading-order (NLO)
heavy-quark corrections to the axial charge $\gaz$ for protons
to couple to the weak neutral current 
\begin{equation}
J_{\mu5}^Z\ 
=\ \smallfrac{1}{2} \biggl\{\,\sum_{q=u,c,t} - \sum_{q=d,s,b}\,\biggr\}\:
        \bar{q}\gamma_\mu\gamma_5q
\label{g}\end{equation}
The calculation is performed by decoupling heavy quarks $h = t,b,c$ 
sequentially, i.e.\ one at a time.  An extension to simultaneous 
decoupling of $t,b,c$ quarks is foreshadowed in our concluding remarks.

The charge $\gaz$ receives contributions from both light $u,d,s$ and
heavy $c,b,t$ quarks,
\begin{equation}
2\gaz = \bigl( \Delta u - \Delta d - \Delta s \bigr) 
       + \bigl( \Delta c - \Delta b + \Delta t \bigr)
\label{g1}\end{equation}
where $\Delta q$ refers to the expectation value
\[ \langle p,s|\,\bar{q}\gamma_\mu\gamma_5q\,|p,s \rangle 
  = 2m_p s_\mu\Delta q \]
for a proton of spin $s_\mu$ and mass $m_p$. It governs parity-violating 
effects due to $Z^0$ exchange at low energies in elastic $\nu p$ 
and $\bar{\nu}p$ scattering \cite{garvey,KM} or in light atoms 
\cite{fortson,ellis}.  A definitive measurement of $\nu p$ elastic scattering
may be possible using the miniBooNE set-up at FNAL \cite{tayloe}.

Once heavy-quark corrections \cite{CWZ,KM,Bass} have been taken into 
account, $\gaz$ is related (modulo the issue of $\delta$--function 
terms at $x=0$ \cite{topology}) to the flavour-singlet axial 
charge, defined scale invariantly and extracted from polarized deep 
inelastic scattering:
\begin{equation}
g_A^{(0)}\bigr|_\mathrm{inv} = 0.2 - 0.35 
\label{inv}\end{equation}
The small value of this quantity has inspired vast experimental and 
theoretical activity to understand the spin structure of the proton 
\cite{spinrev}.  As a result, new experiments are being planned 
to map out the spin-flavour structure of the proton.  These include 
polarized proton-proton collisions at the Relativistic Heavy Ion 
Collider (RHIC) \cite{rhic}, semi-inclusive polarized deep inelastic 
scattering, and polarized $ep$ collider studies \cite{bassdr}.  Full
NLO analyses are essential for a consistent interpretation of these
experiments.

Many techniques for decoupling a single heavy quark are available.  We rely 
on Witten's method \cite{witten}, where the renormalization scheme is 
mass independent and improved Callan-Symanzik equations \cite{Weinberg2}
can be exploited.  In such schemes, the decoupling of heavy particles 
required by the Appelquist-Carrazone theorem \cite{AC} is not manifest. 
However, correct decoupling is ensured by applying the matching 
conditions of Bernreuther and Wetzel \cite{BW}; these relate 
coupling constant, mass and operator normalizations before and after 
the decoupling of a heavy quark.  The advantages of this approach are 
its rigour and the fact that the final results are expressed in terms of
renormalization group (RG) invariants.  These invariants are Witten-style 
running couplings $\widetilde{\alpha}_h$, one for each heavy quark 
$h=t,b,c$, and axial charges for nucleons in the residual theory with 
three light flavours.

We find that, when first $t$, then $b$, and finally $c$ are decoupled 
from (\ref{g1}), the full NLO result is
\begin{align}
2\gaz\, =\, &\bigl(\Delta u - \Delta d - \Delta s\bigr)_\mathrm{inv}
           +\hsp{0.2} {\cal P}\hsp{0.1}\bigl(
               \Delta u + \Delta d + \Delta s\bigr)_\mathrm{inv} 
 \nonumber \\
    &+\, O(m_{t,b,c}^{-1})
\label{g2}\end{align}
where ${\cal P}$ is a polynomial in the running couplings
$\run{h}$,
\begin{align}
{\cal P}\, =\, &\smallfrac{6}{23\pi}\bigl(\run{b}-\run{t}\bigr)
             \Bigl\{1 + \smallfrac{125663}{82800\pi}\run{b}
                      + \smallfrac{6167}{3312\pi}\run{t}
                      - \smallfrac{22}{75\pi}\run{c}  \Bigr\}
\nonumber \\
&\phantom{+\ \Biggl[} - \smallfrac{6}{27\pi} \run{c}
                      - \smallfrac{181}{648 \pi^2}\run{c}^2
                      + O\bigl(\run{t,b,c}^3\bigr)
\label{j3}
\end{align}
and $(\Delta q)_\mathrm{inv}$ denotes the scale-invariant version of 
$\Delta q$ defined in the following way.

Let $\alpha_f = g_f^2/4\pi$ and $\beta_f(\alpha_f)$ be the gluon coupling
and beta function for $\overline{\mbox{\small MS}}$ renormalized 
quantum chromodynamics (QCD) with $f$ flavours and $N_c=3$ colours, 
and let  $\gamma_f(\alpha_f)$ be the gamma function for the singlet current
\begin{equation} 
\bigl(\bar{u}\gamma_\mu\gamma_5u + \bar{d}\gamma_\mu\gamma_5d 
      + \ldots\bigr)_f 
 = \sum_{k=1}^f\bigl(\bar{q}_k\gamma_\mu\gamma_5q_k\bigr)_f 
\label{singlet}\end{equation}
A scale-invariant current $(S_{\mu 5})_f$ is obtained when 
(\ref{singlet}) is multiplied by 
\begin{equation}
E_f(\alpha_f)\, =\, \exp\!\int^{\alpha_f}_0\hsp{-2.5} dx\,
                      \frac{\gamma_f(x)}{\beta_f(x)}
\label{f0}\end{equation}
Up to $O(m_h^{-1})$ corrections, the invariant singlet charge 
(\ref{inv}) is given by
\begin{align}
g_A^{(0)}\bigr|_\mathrm{inv} &=  E_3(\alpha_3) 
       \bigl(\Delta u + \Delta d + \Delta s\bigr)_3   \nonumber \\
&= \bigl(\Delta u + \Delta d + \Delta s\bigr)_\mathrm{inv}
\label{inv2}\end{align}
Flavour-dependent, scale-invariant axial charges\, $\Delta q|_{\rm inv}$ 
such as 
\begin{equation}
\Delta s|_\mathrm{inv} 
 = \smallfrac{1}{3}\Bigl(g_A^{(0)}\bigr|_\mathrm{inv}-g_A^{(8)}\Bigr)
\end{equation}
can then  be obtained from linear combinations of (\ref{inv2}) and
\begin{gather}
g_A^{(3)} = \Delta u - \Delta d  
          = \bigl(\Delta u - \Delta d\bigr)_\mathrm{inv}   \nonumber \\
g_A^{(8)} = \Delta u + \Delta d - 2 \Delta s  
          = \bigl(\Delta u + \Delta d - 2 \Delta s\bigr)_\mathrm{inv}
\end{gather}
Here $g_A^{\scriptscriptstyle (3)} = 1.267 \pm 0.004$ 
is the isotriplet axial charge measured in neutron beta-decay, and 
$g_A^{\scriptscriptstyle (8)} = 0.58 \pm 0.03$ is the
octet charge measured independently in hyperon beta decay. 
Taking $\widetilde{\alpha}_t = 0.1$, $\widetilde{\alpha}_b = 0.2$
and $\widetilde{\alpha}_c = 0.35$ in (\ref{j3}), we find a small 
heavy-quark correction factor ${\cal P}= -0.02$, with LO terms
dominant.

Our results extend and make more precise the well known work of
Collins, Wilczek and Zee \cite{CWZ} and Kaplan and Manohar \cite{KM},
where heavy-quark effective theory was used to estimate $\gaz$ 
in leading order (LO) for sequential decoupling of $t,b$ and $t,b,c$
respectively.  Our analysis is also influenced by a discussion of 
\cite{CWZ} by Chetyrkin and K\"{u}hn \cite{CK}, who considered some
aspects of NLO decoupling of the $t$ quark from the neutral current and 
in particular, the requirement that the result be scale invariant.  
Related work has been done on heavy-quark production 
in polarized deep inelastic scattering using the QCD parton model 
\cite{bbs} and in high-energy polarized $\gamma p$ and $pp$ at NLO 
\cite{bojak}.

The plan of this paper is as follows. 
Section 2 is a brief review of Witten's application of improved 
Callan-Symanzik equations \cite{Weinberg2} to the decoupling of a heavy 
quark in mass-independent renormalization schemes. In Section 3, we 
combine it with matching conditions \cite{BW} to deal with 
next-to-leading-order (NLO) calculations involving axial-vector 
currents.  Section 4 is then a direct derivation of (\ref{j3}) from 
the formula (\ref{g}) for the neutral current. Our concluding remarks in
section 5 indicate the result of extending (\ref{j3}) to simultaneous
decoupling of $t,b,c$ --- done not only for numerical reasons, but also 
to check that the $t,b$ contributions cancel for $m_t=m_b$.

\section{Witten's Method}

In mass-independent schemes such as $\overline{\mbox{\small MS}}$, 
renormalized masses behave like coupling constants.  This key property
is exploited in Witten's method.

Let $\mu$ be the scale used to define dimensional regularization and
renormalization.  Then the $\overline{\mbox{\small MS}}$ scale is
\begin{equation}
\bar{\mu} = \mu\sqrt{4\pi}e^{-\gamma/2}\ ,\ \ \gamma = 0.5772 \ldots
\end{equation}
We choose the same scale $\bar{\mu}$ irrespective of the number of 
flavours $f$ being considered, and so hold $\bar{\mu}$ fixed as the 
heavy quarks (masses $m_h$) decouple: 
\[  F\, \to\, f\ \mbox{flavours},\,\ m_h \to \infty  \]
Also held fixed in this limit are the coupling $\alpha_f$ and light-quark 
masses ${m_\ell}_f$ of the \emph{residual} $f$-flavour theory, and all 
momenta $\mathbf{p}$.  Feynman diagrams for amplitudes
\begin{equation}
{\cal A}_F = 
{\cal A}_F\bigl(\mathbf{p}, \bar{\mu}, \alpha_F, {m_\ell}_F, m_h\bigr) 
\end{equation}
give rise to power series in $\smash{m_h^{-1}}$ modified by polynomials 
in $\ln(m_h/\bar{\mu})$.  We consider just the leading power
$\widetilde{\cal A}_F$: 
\begin{equation}  
{\cal A}_F = \widetilde{\cal A}_F\{1 + O(1/m_h)\}  
\label{c} 
\end{equation}

As $m_h$ tends to infinity, logarithms in $\widetilde{\cal A}_F$ can be 
produced by any 1PI (one-particle irreducible) subgraph which 
contains at least one heavy-quark propagator and whose divergence by 
power counting is at least logarithmic.  The effect is equivalent to
shrinking all contributing 1PI parts of each diagram to a point. This 
means \cite{AC} that the $F$-flavour amplitudes $\widetilde{\cal A}_F$ 
are the same as amplitudes ${\cal A}_f$ in the residual $f$-flavour theory, 
apart from $m_h$-dependent renormalizations of the coupling constant, 
light masses, and amplitudes: 
\begin{align}
\widetilde{\cal A}_F\bigl(\mathbf{p}&, \bar{\mu},\alpha_F,{m_\ell}_F,m_h\bigr) 
  \nonumber \\
&=\, \sum_{{\cal A}'}{\cal Z}_{\!\cal AA'}(\alpha_F, m_h/\bar{\mu})
{\cal A}'_f\bigl({\bf p},\bar{\mu},\alpha_f,{m_\ell}_f\bigr) 
\label{a} \\
\alpha_f =\ &\alpha_f(\alpha_F, m_h/\bar{\mu}) \hsp{1},\hsp{1}
{m_\ell}_f = {m_\ell}_FD(\alpha_F, m_h/\bar{\mu}) 
\label{a1}\end{align}
Eventually, we will have to invert (\ref{a1}), i.e.\ use $\alpha_f$ and
${m_\ell}_f$ as dependent variables instead of $\alpha_F$ and
${m_\ell}_F$, because we hold $\alpha_f$ and ${m_\ell}_f$ fixed as
$m_h \to \infty$.

For any number of flavours $f$ (including $F$), let
\begin{equation}
{\cal D}_f\, =\, \bar{\mu}\smallfrac{\del\ }{\del\bar{\mu}}
            + \beta_f(\alpha_f)\smallfrac{\del\ \ }{\del\alpha_f}
        + \delta_f(\alpha_f)\sum_{k=1}^f {m_k}_{\hsp{-0.2}f}
               \smallfrac{\del\hsp{4.6}}{\del {m_k}_{\hsp{-0.2}f}}
\label{b7}
\end{equation}
be the corresponding Callan-Symanzik operator.  Then the amplitude 
${\cal A}_F$ and hence its leading power $\widetilde{\cal A}_F$ both 
satisfy an $F$-flavour improved Callan-Symanzik equation:
\begin{equation} 
\bigl\{{\cal D}_F + \gamma_F(\alpha_F)\bigr\}\widetilde{\cal A}_F\, =\, 0   
\label{d} 
\end{equation}
In general, both  $\gamma_F$ and 
${\cal Z} = \bigl({\cal Z}_{\!\cal AA'}\bigr)$ are matrices. \vspace{0.5mm}

If we substitute (\ref{a}) in (\ref{d}) and change variables,
\begin{equation}
{\cal D}_F\, =\, \bar{\mu}\smallfrac{\del\ }{\del\bar{\mu}}
   + \bigl({\cal D}_F\alpha_f\bigr)\smallfrac{\del\ \ }{\del\alpha_f}
   + \sum_{k=1}^f\bigl({\cal D}_F{m_k}_{\hsp{-0.2}f}\bigr)
              \smallfrac{\del\hsp{4.6}}{\del {m_k}_{\hsp{-0.2}f}} 
\end{equation}
the result is an improved Callan-Symanzik equation for each residual 
amplitude,
\begin{equation} 
\bigl\{{\cal D}_f + \gamma_f(\alpha_f)\bigr\}{\cal A}_f\, =\, 0
\end{equation}
where the functions \cite{witten,BW}
\begin{eqnarray}
\beta_f(\alpha_f) &=& {\cal D}_F\alpha_f  
\label{e1} \\
\delta_f(\alpha_f) &=& {\cal D}_F\ln m_\ell 
\label{e2} \\
\gamma_f(\alpha_f) &=& 
     {\cal Z}^{-1}\bigl(\gamma_F(\alpha_F) + {\cal D}_F\bigr){\cal Z} 
\label{e3} 
\end{eqnarray}
depend on $\alpha_f$ \emph{alone}.  The lack of $m_\ell$ dependence
of the renormalization factors in (\ref{a}) and (\ref{a1}) ensures
mass-independent renormalization for the residual theory.

Although these equations hold for any $f<F$, their practical
application is straightforward only when heavy quarks are decoupled 
one at a time.  So we set $F = f\!+\!1$, where just one quark $h$ 
is heavy.  Then it is convenient to introduce a running 
coupling \cite{witten}
\begin{equation} 
\run{h} = \run{h}\bigl(\alpha_F, \ln(m_h/\bar{\mu})\bigr)
\end{equation} 
associated with the $\overline{\mbox{\small MS}}_F$ renormalized mass $m_h$:
\begin{equation} 
\ln(m_h/\bar{\mu})\ 
=\ \int^{\run{h}}_{\alpha_F}\! dx\,\bigl(1-\delta_F(x)\bigr)/\beta_F(x)
\label{f3} 
\end{equation}
It satisfies the constraints
\begin{equation} 
\run{h}(\alpha_F,0) = \alpha_F \hsp{3},\hsp{3}  
\run{h}(\alpha_F,\infty) = 0
\end{equation}
the latter being a consequence of the asymptotic freedom of the $F$ 
flavour theory ($F \leqslant 16$). Also, eqs.~(\ref{b7}), (\ref{e1}) 
and (\ref{f3}) imply that $\run{h}$ is renormalization group (RG) 
invariant:  \vspace{-1mm}
\begin{equation} 
{\cal D}_F \run{h} = 0
\end{equation}

Witten's solution of (\ref{e3}) for the matrix ${\cal Z}$ is 
\begin{align}
{\cal Z}(\alpha_F,m_h/\mu)\, =\, 
 &\exp\biggl\{\int^{\run{h}}_{\alpha_F}\!\!dx\, 
  \frac{\gamma_F(x)}{\beta_F(x)}\biggr\}_\mathrm{ord}{\cal Z}(\run{h},1)
\nonumber \\  &\times
 \exp-\biggl\{\int^{\alpha_f(\run{h},1)}_{\alpha_f}\!\!\! dx\,
         \frac{\gamma_f(x)}{\beta_f(x)}\biggr\}_\mathrm{ord}
\label{g4}\end{align}
where ``ord'' indicates $x$-ordering of matrix integrands in the 
exponentials.  Note that it is the \emph{relative} scaling 
between the initial and residual theories which matters.

For our NLO calculation, we need the formulas
\begin{eqnarray}
\beta_f(x)
&=&
 - \smallfrac{x^2}{3\pi}\Bigl(\footfrac{33}{2}\!-\!f\Bigr)
    - \smallfrac{x^3}{12\pi^2}(153\!-\!19f) + O(x^4) \ \ \
\nonumber \\
\gamma_f(x)
&=& 
  \smallfrac{x^2}{\pi^2}f + \smallfrac{x^3}{36\pi^3}
            \bigl(177\!-\!2f\bigr)f + O(x^4) \ \ \ 
\nonumber \\
\delta_f(x) 
&=&
  - \smallfrac{2x}{\pi} + O(x^2)\ \ \ 
\label{e1a}\end{eqnarray}
where $\gamma_f$ refers to the $f$-flavour singlet current
(\ref{singlet}) and includes the three-loop term found by Larin 
\cite{Larin} and Chetyrkin and K\"{u}hn \cite{CK}.

\section{Matching procedure}

Our task is to evaluate to NLO accuracy the quantities 
$\run{h}$, $\alpha_f(\run{h},1)$ and ${\cal Z}(\run{h},1)$ 
in (\ref{g4}), such that the answers depend on $\alpha_f$ and
not $\alpha_F$.

Bernreuther and Wetzel \cite{BW} applied the Appelquist-Carrazone 
decoupling theorem \cite{AC} to the gluon coupling constant 
$\alpha^{\mbox{\tiny MO}}_Q$ renormalized at space-like momentum $Q$,
\begin{equation}
\alpha^{\mbox{\tiny MO}}_Q\bigr|_{\mathrm{with}\ h}\;
   =\; \alpha^{\mbox{\tiny MO}}_Q\bigr|_{\mathrm{no}\ h}\,
          +\,  O(m^{-1}_h)
\end{equation}
and compared calculations of $\alpha^{\mbox{\tiny MO}}_Q$  in the 
$F=f\!+\!1$ and $f$ flavour $\overline{\mbox{\small MS}}$ theories. 
This reduces to a determination of the leading power of the one-$h$-loop
$\overline{\mbox{\small MS}}_F$ gluon self-energy. The result is a
matching condition
\begin{equation}
\alpha_F^{-1} - \alpha_f^{-1}\,
 =\, C_\mathrm{LO}\ln\smallfrac{m_h}{\bar{\mu}}
     + C_\mathrm{NLO} + O(\alpha_f, m_h^{-1})
\label{a7}\end{equation}
with $\alpha_f$-independent LO and NLO coefficients given by
\begin{equation}
C_\mathrm{LO} = \smallfrac{1}{3\pi} \hsp{3},\hsp{3} C_\mathrm{NLO} = 0
\label{match}\end{equation}
As a result, we find:
\begin{equation}
\alpha_f(\run{h},0) 
= \run{h} + O(\run{h}^3) \underset{\mathrm{NLO}}{=} \run{h} 
\end{equation}

Bernreuther and Wetzel showed that it is possible to deduce \emph{all} 
LO and NLO terms in (\ref{a7}) from (\ref{match}) and $\beta_f$ and 
$\delta_f$ in (\ref{e1a}). We have done the calculation explicitly:
\begin{align}
\alpha_{f+1}^{-1} &\underset{\mathrm{NLO}}{=}
\alpha_f^{-1} + \smallfrac{1}{3\pi}\ln\smallfrac{m_h}{\bar{\mu}} +
c_f\ln\Bigl[1 + \smallfrac{\alpha_f}{3\pi}\ln\smallfrac{m_h}{\bar{\mu}}\Bigr]
\nonumber \\
&\phantom{\underset{\mathrm{NLO}}{=} \alpha_f^{-1}\ }
+ d_f\ln\Bigl[1 + \smallfrac{\alpha_f}{3\pi}\Bigl(\smallfrac{33}{2}-f\Bigr)
      \ln\smallfrac{m_h}{\bar{\mu}}\Bigr]
 \nonumber \\
c_f &=  \smallfrac{142-19f}{2\pi(31\!-\!2f)} \hsp{2},\hsp{2}
d_f =  \smallfrac{57+16f}{2\pi(33\!-\!2f)(31\!-\!2f)}
\end{align}
{}From (\ref{f3}), we have also found $\run{h}$ in NLO,
\begin{align}
\run{h}^{-1} \underset{\mathrm{NLO}}{=}\
 &\alpha_f^{-1} + \smallfrac{1}{3\pi}\Bigl(\smallfrac{33}{2}-f\Bigr)
                           \ln\smallfrac{\bar{m}_h}{\bar{\mu}}
\nonumber \\
 &\!+ \smallfrac{153-19f}{2\pi(33\!-\!2f)} 
  \ln\Bigl[1 + \smallfrac{\alpha_f}{3\pi}\Bigl(\smallfrac{33}{2}-f\Bigr)
      \ln\smallfrac{m_h}{\bar{\mu}}\Bigr]
\end{align}
where $\bar{m}_h$ is Witten's RG invariant mass:
\begin{equation}
\bar{m}_h = m_h\exp\int^{\run{h}}_{\alpha_F}\!\!dx\, \delta_F(x)/\beta_F(x)
\end{equation}
If desired, $\ln(\bar{m}_h/\bar{\mu})$ can be eliminated by substituting
\begin{equation}
\ln\smallfrac{\bar{m}_h}{\bar{\mu}}\,
 \underset{\mbox{\scriptsize LO}}{=}\,
   \ln\smallfrac{m_h}{\bar{\mu}} - \smallfrac{12}{31\!-\!2f}\ln \Bigl[
      1 + \smallfrac{\alpha_f}{3\pi}\Bigl(\smallfrac{31}{2}\!-\!f\Bigr)
               \ln\smallfrac{m_h}{\bar{\mu}}\Bigr]
\end{equation}
Therefore the asymptotic formula for $\run{h}$ as $m_h \to \infty$ is
\begin{gather}
\widetilde{\alpha}_h\ \sim\ 3\pi \!\Bigm/\! \Bigl\{
  \Bigl(\footfrac{33}{2}-f\Bigr)\ln\smallfrac{m_h}{\bar{\mu}}
 + k_f\ln\ln\smallfrac{m_h}{\bar{\mu}}
 + O(1)\Bigr\}  \nonumber \\
k_f = \smallfrac{3(153-19f)}{2(33-2f)} - \smallfrac{6(33-2f)}{31-2f}
\end{gather}

To find the matrix ${\cal Z}(\run{h},1)$ in NLO, we need a matching
condition for the $\overline{\mbox{\small MS}}$ amplitude 
$\Gamma_{\mu 5}$ for $\bar{h}\gamma_\mu\gamma_5 h$ to couple to
a light quark $\ell$. We have calculated the leading power due to
the two-loop diagram
\setlength{\unitlength}{0.45mm}
\begin{picture}(16,3)(0,1)
\newcommand{\glueA}%
{\curve(0,0, 0.6666,0.8, 1.3333,0)\curve(1.3333,0, 2,-0.8, 2.6666,0)
\curve(2.6666,0, 3.3333,0.8, 4,0)\curve(4,0, 4.6666,-0.8, 5.3333,0)
\curve(5.3333,0, 6,0.8, 6.6666,0)\curve(6.6666,0, 7.3333,-0.8, 8,0)}
\curve(0,3, 4,6)\curve(0,3, 4,0)\curve(4,0, 4,6)
\newcommand{\glueB}%
{\curve(0,0, 0.6666,-0.8, 1.3333,0)\curve(1.3333,0, 2,0.8, 2.6666,0)
\curve(2.6666,0, 3.3333,-0.8, 4,0)\curve(4,0, 4.6666,0.8, 5.3333,0)
\curve(5.3333,0, 6,-0.8, 6.6666,0)\curve(6.6666,0, 7.3333,0.8, 8,0)}
\curve(0,3, 4,6)\curve(0,3, 4,0)\curve(4,0, 4,6)
\put(4,0){\glueB}\put(4,6){\glueA}
\curve(12,0, 12,6)\curve(12,0, 16,-1)\curve(12,6, 16,7)
\end{picture}\ :
\begin{equation}
\Gamma_{\mu 5}\,
 =\, \Bigl(\smallfrac{\alpha_F}{\pi}\Bigr)^2 \gamma_\mu\gamma_5
     \Bigl(\ln\smallfrac{m_h}{\bar{\mu}} + \smallfrac{1}{8}\Bigr)\,
     +\, O\bigl(\alpha_F^3,m_h^{-1}\bigr)
\label{f4}\end{equation}
Consequently,  there is a NLO term $\run{h}^2/8\pi^2$ in 
${\cal Z}(\run{h},1)$ for $\bar{h}\gamma_\mu\gamma_5 h$ to produce 
$\bar{\ell}\gamma_\mu\gamma_5\ell$ as $m_h \to \infty$.

\section{Heavy Quarks decoupled from $\bm{J_{\mu 5}^{Z}}$}

Let us adopt the shorthand notation $q_f$ for 
$\overline{\mbox{\small MS}}$ currents
$(\bar{q}\gamma_\mu\gamma_5 q)_f$ in
the $f$-flavour theory, e.g.\ the neutral current 
$J_{\mu 5}^{\mbox{$\scriptscriptstyle (Z)$}}$ and
the scale-invariant singlet current $(S_{\mu 5})_f$:
\begin{align} 
J^{\rm Z} &= \footfrac{1}{2}\bigl(t-b+c-s+u-d\bigr)_6   \\
      S_f &= E_f(\alpha_f)(u + d + s + \ldots)_f      \label{scale}
\end{align}

We begin by decoupling the $t$ quark. Because of
\begin{equation}
(c-s+u-d)_6 = (c-s+u-d)_5 + O(1/m_t)
\label{nonsing}\end{equation}
we see that (\ref{g4}) is non-trivial only for
\begin{align}
(t-b)_6 =\ &{\cal Z}_{6 \to 5}(u+d+s+c+b)_5   \nonumber \\
          &+ \footfrac{1}{5}(u+d+s+c-4b)_5 + O(1/m_t)
\label{e8}
\end{align}
Since $(t-b)_6$ is scale invariant, we have $\gamma_F = 0$ in (\ref{g4}): 
\begin{equation}
 {\cal Z}_{6 \to 5}(\alpha_6,m_t/\bar{\mu})\
\underset{\mbox{\scriptsize NLO}}{=}\
 {\cal Z}_{6 \to 5}(\run{t},1)
     \exp -\!\int^{\run{t}}_{\alpha_5}\hsp{-2.5} dx\,
         \frac{\gamma_5(x)}{\beta_5(x)}
\label{e9} \end{equation}
The operator matching condition (\ref{f4}) corresponds to
\begin{equation}
t_6\, =\, \smallfrac{\alpha_6^2}{\pi^2}
          \Bigl(\ln\smallfrac{m_t}{\bar{\mu}} + \smallfrac{1}{8}\Bigr)
          (u+d+s+c+b)_5  
          +\, O(\alpha_6^3,m_t^{-1})
\end{equation}
and so we conclude:
\begin{equation}
{\cal Z}_{6 \to 5}(\widetilde{\alpha}_t,1)\,
=\, - \smallfrac{1}{5} + \smallfrac{1}{8\pi^2}\widetilde{\alpha}_t^2
                       + O(\widetilde{\alpha}_t^3)
\end{equation}

Eq.~(\ref{e9}) is to be expanded about $\run{t} \sim 0$ with $\alpha_5$
held fixed.  In that limit, the exponential tends to the constant
factor $E_5(\alpha_5)$ of (\ref{f0}). This factor combines with the
singlet current in (\ref{e8}) to form the scale-invariant operator 
$S_5$, as required by RG$_{f=5}$ invariance.  The full NLO result is 
then obtained by writing
\begin{align}
(t-b)_6 \underset{\mbox{\scriptsize NLO}}{=}\
   &{\cal Z}_{6 \to 5}(\run{t},1)
     \exp\biggl\{-\!\int^{\run{t}}_0 \hsp{-2.5}
          dx\,\frac{\gamma_5(x)}{\beta_5(x)}\biggr\}S_5
\nonumber \\
&+ \footfrac{1}{5}(u+d+s+c-4b)_5
\end{align}
and expanding in $\run{t}$, keeping all quadratic terms:
\begin{align}
(t-b)_6 =\ 
&\Bigl\{- \smallfrac{1}{5}
      - \smallfrac{6}{23}\smallfrac{\run{t}}{\pi}
  \Bigl(1 + \smallfrac{6167}{3312}\smallfrac{\run{t}}{\pi}\Bigr)
      + O\bigl(\run{t}^3\bigr)\Bigr\} S_5
\nonumber \\
&+ \footfrac{1}{5}(u+d+s+c-4b)_5 + O(1/m_t)
\label{f6}
\end{align}

Next we decouple the $b$ quark. Here, it is natural to define five-flavour
quantities $\mbox{$\run{b}$}_5$ and $\mbox{$\bar{m}_b$}_5$
analogous to the six-flavour running coupling $\run{t}$
and mass $\bar{m}_t$ for the top quark:
\begin{equation}
\ln\smallfrac{{m_b}_5}{\bar{\mu}}\,
=\, \int_{\alpha_5}^{\mbox{$\scriptstyle \widetilde{\alpha}_b$}_5}
      \hsp{-2.5}dx\,\smallfrac{1-\delta_5(x)}{\beta_5(x)}
\hsp{3},\hsp{3}
\ln\smallfrac{\mbox{\small $\bar{m}_b$}_5}{{m_b}_5}
=\, \int_{\alpha_5}^{\mbox{$\scriptstyle \widetilde{\alpha}_b$}_5}
      \hsp{-2.5}dx\,\smallfrac{\delta_5(x)}{\beta_5(x)}
\end{equation}
Eqs.~(\ref{e1}) and (\ref{e2}) imply that 
$\mbox{$\run{b}$}_5$ and $\mbox{$\bar{m}_b$}_5$ are both
RG$_{f=5}$ and RG$_{f=6}$ invariant
\begin{equation}
{\cal D}_5\mbox{$\run{b}$}_5 = 0 = {\cal D}_6\mbox{$\run{b}$}_5
\hsp{3},\hsp{3}
{\cal D}_5\mbox{$\bar{m}_b$}_5 = 0 = {\cal D}_6\mbox{$\bar{m}_b$}_5
\end{equation}
and hence physically significant in the original six-flavour theory.
So we write $\run{b}$ and $\bar{m}_b$ for
$\mbox{$\run{b}$}_5$ and $\mbox{$\bar{m}_b$}_5$.

Consider decoupling the $b$ quark from (\ref{f6}). The NLO matching 
condition (\ref{f4}) becomes
\begin{equation}
b_5\, =\, \smallfrac{\alpha_5^2}{\pi^2}
     \Bigl(\ln\smallfrac{\mbox{\small $\bar{m}_b$}_5}{\bar{\mu}}
    + \smallfrac{1}{8}\Bigr)(u+d+s+c)_4 
    + O(\alpha_5^3,\mbox{$m_b$}_5^{-1})
\end{equation}
so the non-singlet current in (\ref{f6}) can be written
\begin{align}
&(u+d+s+c-4b)_5  \nonumber \\
&\phantom{(u+d\,}
=\, \Bigl\{1 - \smallfrac{\run{b}^2}{2\pi^2}\Bigr\}E_4^{-1}(\run{b})S_4 
      + O(\run{b}^3,\mbox{$m_b$}_5^{-1})
\label{i4}
\end{align}
For the singlet current $S_5$ in (\ref{f6}). we find
\begin{equation}
S_5 = E_5(\widetilde{\alpha}_b)
     \Bigl\{1 + \smallfrac{\widetilde{\alpha}_b^2}{8\pi^2}
      \Bigr\}E_4^{-1}(\widetilde{\alpha}_b)S_4
    + O(\run{b}^3,\mbox{$m_b$}_5^{-1})
\label{i7} \end{equation}
taking into account the definitions (\ref{f0}) and (\ref{scale}). 
Then we expand (\ref{i4}) and (\ref{i7}) in $\run{b}$, keeping quadratic
terms:
\begin{align}
(t-b)_6\, =\,
 &\smallfrac{6}{23\pi}\bigl(\widetilde{\alpha}_b-\widetilde{\alpha}_t\bigr)
 \Bigl\{1 + \smallfrac{125663}{82800\pi}\widetilde{\alpha}_b
          + \smallfrac{6167}{3312\pi}\widetilde{\alpha}_t\Bigr\}S_4
\nonumber \\[1mm]
 &+\, O(\widetilde{\alpha}_{t,b}^3,m_{t,b}^{-1})
\label{j4}
\end{align}

The same technique can be applied to decouple the $c$ quark from $S_4$
in (\ref{j4}) and\, $(c-s+u-d)_4$\, (the result of decoupling $b$ from 
(\ref{nonsing})). That yields the final results (\ref{g2}) and (\ref{j3}) 
given in the Introduction.

\section{Remarks}

Our results depend on two key features:
\begin{enumerate}
\item Like previous workers in this area, we decouple heavy quarks 
      sequentially, i.e.\ one at a time.
\item Our running couplings $\run{t}$, $\run{b}$ and $\run{c}$, which
      correspond to Witten's prescription \cite{witten}, are all
      renormalization group invariant.
\end{enumerate}      

The restriction to sequential decoupling is numerically reasonable
for the $t$ quark, but dubious for the $b$ and $c$ quarks, because it 
amounts to an assumption that $\ln(m_c/\bar{\mu})$
is negligible compared with $\ln(m_b/\bar{\mu})$.  This inhibits
detailed comparison of NLO results with data, which
ought to be carried out with NLO accuracy \cite{remarks}.

There is also a theoretical issue here: one would like to check that, 
in the limit $m_t = m_b$, the $t$ and $b$ contributions cancel. 
However, that is outside the region of validity\,  
$\ln(m_t/\bar{\mu}) \gg \ln(m_b/\bar{\mu})$\,
for sequential decoupling.

For these reasons, we have extended our analysis to the case of 
simultaneous decoupling, where the mass logarithms are allowed to grow
large together:
\[ \ln(m_c/\bar{\mu}) \sim \ln(m_b/\bar{\mu}) \sim \ln(m_t/\bar{\mu})\,
\to\, \mbox{large}  \]
This requires a considerable theoretical development of matching
conditions and the renormalization group, which we will present 
separately.  It involves the construction of running couplings
$\alpha_t$, $\alpha_b$, $\alpha_c$ with the following properties:
\begin{enumerate}
\item They are renormalization group invariant.
\item They are defined for $m_t \geqslant m_b \geqslant m_c$,
      and can have a non-trivial dependence on more than one 
      heavy-quark mass.  
\item In the special case of sequential decoupling, they agree with 
      $\run{t}$, $\run{b}$ and $\run{c}$ to NLO.
\item For the case of equal masses, they coincide, e.g.
\begin{equation}  
\alpha_t = \alpha_b\ \ \mbox{for}\ \ m_t = m_b
\label{equal}\end{equation}
\end{enumerate}

Then we find that the result for the simultaneous decoupling of the 
$t,b,c$ quarks from the neutral current is of the same form (\ref{g2})
as the sequential answer, but with the sequential running couplings 
in (\ref{j3}) replaced by our simultaneous couplings $\alpha_t$, 
$\alpha_b$, and $\alpha_c$:
\begin{align}
{\cal P}\, =\, &\smallfrac{6}{23\pi}\bigl(\alpha_b-\alpha_t\bigr)
             \Bigl\{1 + \smallfrac{125663}{82800\pi}\alpha_b
                      + \smallfrac{6167}{3312\pi}\alpha_t
                      - \smallfrac{22}{75\pi}\alpha_c\Bigr\}
\nonumber \\
&\phantom{+\ \Biggl[} - \smallfrac{6}{27\pi}\alpha_c
                      - \smallfrac{181}{648 \pi^2}\alpha_c^2
                      + O\bigl(\alpha_{t,b,c}^3\bigr)
\end{align}
Notice the factorization of the terms depending on $\alpha_t$ and 
$\alpha_b$.  Given (\ref{equal}), the factor $\alpha_b-\alpha_t$ 
ensures that all contributions from $b$ and $t$ quarks cancel 
(as they should) for $m_t=m_b$.

\acknowledgments
This work was supported by the Australian Research Council and the
Austrian FWF. FMS is supported by contract number PV-IFT/005.  
RJC thanks Professor Wojtek Zakrzewski for his hospitality at Durham.
SDB thanks Professor Dietmar Kuhn and the HEP group for their hospitality at
Innsbruck.


\begin{thebibliography}{99}
\bibitem{remarks}
This includes matching conditions for the $b$ and $c$ masses, to be 
discussed elsewhere.
\end{thebibliography}
\end{document}